# Broadband-tunable LP$_{01}$ mode frequency shifting by Raman coherence waves in H$_2$-filled hollow-core PCF

S. T. BAUERSCHMIDT,* D. NOVOA, A. ABDOLVAND, AND P. ST.J. RUSSELL

*Max Planck Institute for the Science of Light, Guenther-Scharowsky Strasse 1, 91058 Erlangen, Germany*

*Corresponding author: sebastian.bauerschmidt@mpl.mpg.de

**When a laser pump beam of sufficient intensity is incident on a Raman-active medium such as hydrogen gas, a strong Stokes signal, red-shifted by the Raman transition frequency $\Omega_R$, is generated. This is accompanied by the creation of a "coherence wave" of synchronized molecular oscillations with wavevector $\Delta\beta$ determined by the optical dispersion. Within its lifetime, this coherence wave can be used to shift by $\Omega_R$ the frequency of a third "mixing" signal, provided phase-matching is satisfied, i.e., $\Delta\beta$ is matched. Conventionally this can be arranged using non-collinear beams or higher-order waveguide modes. Here we report collinear phase-matched frequency shifting of an arbitrary mixing signal using only the fundamental LP$_{01}$ modes of a hydrogen-filled hollow-core PCF. This is made possible by the S-shaped dispersion curve that occurs around the pressure-tunable zero dispersion point. Phase-matched frequency shifting by 125 THz is possible from the UV to the near-IR. Long interaction lengths and tight modal confinement reduce the peak intensities required, allowing conversion efficiencies in excess of 70%. The system is of great interest in coherent anti-Stokes Raman spectroscopy and for wavelength-conversion of broadband laser sources.**

Coherence waves (C$_w$'s) of collective molecular oscillation, created by stimulated Raman scattering (SRS) [1-5], are useful in the synthesis of ultrashort pulses [6] and for efficient frequency conversion [3-5]. They arise through the beating between two intense quasi-monochromatic laser beams whose frequencies differ by the Raman transition frequency $\Omega_R$. The resulting C$_w$ takes the form of a travelling refractive index grating. Under phase-matched conditions, an arbitrarily weak mixing signal at a different wavelength can be scattered off this C$_w$, which is coherently amplified when the mixing signal is down-shifted by $\Omega_R$, and absorbed when the mixing signal is up-shifted by $\Omega_R$ [2-5].

Phase-matching can be achieved using non-collinear beams in bulk materials [7, 8] or higher-order guided modes in optical fibers [5, 9]. Broadband-guiding kagomé-style hollow-core photonic crystal fiber (kagomé-PCF) is particularly attractive since the dispersion can be precisely controlled by varying the gas pressure [10]. In addition, tight modal confinement and long light-gas interaction lengths dramatically reduce the peak laser intensities required for efficient excitation of C$_w$'s.

In this Letter we report how a C$_w$, excited by an LP$_{01}$ pump pulse in hydrogen-filled kagomé-PCF, can be used, within its coherence lifetime, for efficient phase-matched frequency-shifting of an LP$_{01}$ mixing pulse at an entirely different wavelength. The process is made possible by the unique S-shaped dispersion curve that forms around the pressure-tunable zero dispersion point (ZDP) in kagomé-PCF [10].

To illustrate the concept, the dispersion curve in the vicinity of a ZDP at wavevector $\beta_0$ and frequency $\omega_0$ is drawn schematically in Fig. 1, assuming no dispersion terms higher than third-order. A C$_w$ is first "written" by pump and Stokes waves, labeled by $W_0$ and $W_{-1}$ on the diagram. The $\omega$-$\beta$ "four vector" of the resulting C$_w$ is marked by the blue dashed line. Once created, this C$_w$ can be used for fully phase-matched frequency up-conversion to point $M_1$ of an optical mixing signal at point $M_0$ on the opposite side of the ZDP. It may also be used to seed frequency down-conversion to $M_0$ of a signal at $M_1$, during which process it will be amplified. This procedure will of course also work the other way around, i.e., with writing signals on the low-frequency side of the ZDP and mixing signals on the high-frequency side.

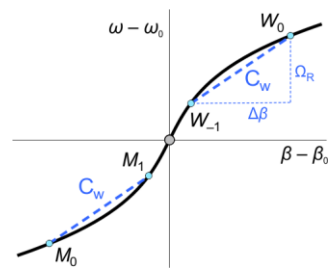

Fig. 1. (a) Sketch of the dispersion of a gas-filled kagomé-PCF in the vicinity of the ZDP (the grey dot at $\omega = \omega_0$ and $\beta = \beta_0$), assuming no dispersion terms higher than third-order. The blue dashed lines indicate the coherence wave C$_w$ excited by the writing signals $W_0$ and $W_{-1}$. C$_w$ is a four-vector with frequency $\Omega_R$ and wavevector $\Delta\beta$ given by the difference between the wavevectors of the writing signals. It can be used either to up-shift the frequency of a mixing signal $M_0$ placed at the position symmetric to $W_0$ on the opposite side of the ZDP or to seed down-conversion of a signal placed at $M_1$.

In general, higher-order dispersion will increasingly distort the perfect S-shape as one moves further away from the ZDP. Nevertheless, as we shall see, it is always possible to find two distinct pairs of points, albeit asymmetrically placed about the ZDP, that are connected by the identical C$_w$. If the frequency of the mixing

beam is not precisely correct, however, the up-conversion process will be dephased at a rate given by:

$$\vartheta = \Delta\beta - (\beta_{M_1} - \beta_{M_0}) \quad (1)$$

where the full dispersion relation of the LP$_{01}$ mode must be taken into account. Its propagation constant $\beta_{01}$ can be approximated to good accuracy by:

$$\beta_{01} = \sqrt{k_0^2 n_{gas}^2(p,\lambda) - u^2/a^2(\lambda)} \quad (2)$$

where $k_0 = 2\pi/\lambda$ is the vacuum wavevector, $n_{gas}$ is the refractive index of the filling gas, $p$ is the gas pressure, $u = 2.405$ for the fundamental core mode and $a(\lambda) = a_{AP}(1 + s\lambda^2/(a_{AP}d))$, where $a_{AP}$ is the area-preserving core radius, $s = 0.08$ is an empirical dimensionless parameter and $d$ is the core-wall thickness [11]. This dispersion relation is plotted in Fig. 2(a) at three different pressures for the fiber used in the experiments ($a_{AP} = 23.4$ μm and $d = 97$ nm). In order to magnify the very flat S-shape, we have subtracted off the linear dispersion by plotting frequency versus ($\beta_{ref} - \beta$), where $\beta_{ref}$ is a pressure-dependent linear function of frequency chosen such that ($\beta_{ref} - \beta$) is zero at 800 THz.

The wavelength of the ZDP shifts from ~400 nm to ~1.1 μm as the pressure increases from 1 to 100 bar [Fig. 2(b)].

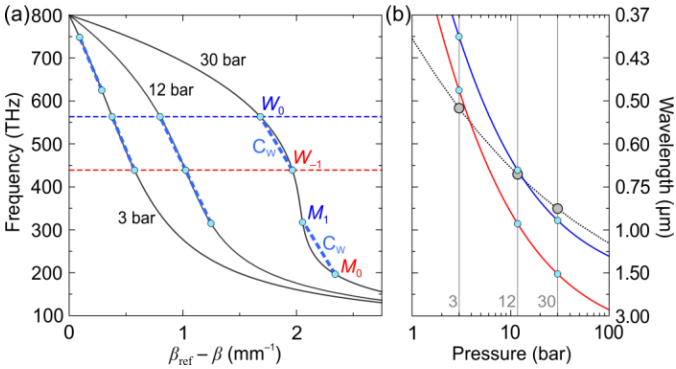

Fig. 2. (a) Dispersion curves of the LP$_{01}$ mode for pressures of 3, 12, and 30 bar (see text). The notation is the same as in Fig. 1, and for clarity only the 30 bar case is fully labeled. (b) Phase-matched frequency pairs (solid lines) and ZDPs (dotted line) as a function of pressure for a pump frequency ($W_0$) of 563 THz (532 nm). The ZDPs for the three pressures in (a) are marked with grey dots.

The mixing frequencies of perfectly phase-matched pairs can be accurately calculated by solving Eqs. 1 and 2. Figure 2(a) shows the $C_w$'s created at three different pressures by beating a pump signal $W_0$ at 532 nm (563 THz) with a Stokes signal $W_{-1}$ at 685 nm (438 THz). The presence of higher-order dispersion shifts the positions of the two perfectly phase-matched pairs asymmetrically about the ZDP. At 30 bar both pump ($W_0$) and Stokes ($W_{-1}$) signals lie in the normal dispersion region ($\omega > \omega_0$), and the phase-matched mixing pair ($M_1$, $M_0$) is at (322, 197) THz in the anomalous dispersion region. At 3 bar, on the other hand, the $C_w$ is excited in the anomalous dispersion region and the mixing pair is phase-matched at (749, 624) THz in the normal dispersion region.

A special situation occurs at 12 bar, when the lower frequency of the upper pair and the upper frequency of the lower pair coincide; this causes a strong second Stokes ($W_{-2}$) signal to appear when pumping with $W_0$ (see below). Overall, Fig. 2 shows that, for a fixed $W_0$ frequency of 563 THz, the frequency of $M_1$ for perfect phase-matching can be tuned from the UV to the near IR simply by changing the pressure.

To confirm these predictions experimentally, we developed a set-up for exciting and probing $C_w$'s under different conditions [see Fig. 3]. The output of a linearly-polarized 1064 nm laser, delivering ns pulses, was split into two parts. The first part was used to generate the $W_0$ signal by frequency-doubling to 532 nm in a KTP crystal. The second was spectrally broadened to ~350 nm in a 4-m-long solid-core PCF, and used as the $M_0$ mixing signal. The pulse durations were 3.2 ns for $W_0$ and 2.0 ns for $M_0$. Both pulses were then recombined and launched into the LP$_{01}$ mode of a 1 m long gas-filled kagomé-PCF. The signals exiting the fiber endface were monitored using an optical spectrum analyzer and imaged using a CCD camera. The intensity of the mixing beam was kept low so as to avoid the onset of SRS, which would influence the $C_w$. The interaction was found to be strongest when the mixing pulse was delayed by ~1 ns from the pump pulse.

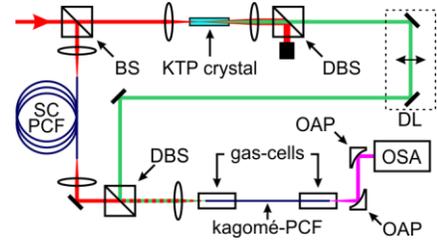

Fig. 3. Schematic of the experimental setup. BS: beam-splitter; DBS: dichroic beam-splitter; OAP: off-axis parabolic mirror; SC-PCF: solid-core photonic crystal fiber; DL: delay line; OSA: optical spectrum analyzer.

The vibrational Raman gain of hydrogen saturates for pressures above ~10 bar [12]. In this regime the behavior of the system is dominated by changes in the phase-matching conditions. Below 10 bar both the phase-matching conditions and the Raman gain are pressure dependent and play a role.

We recorded the spectra generated when scanning the pressure from 5 to 40 bar, for two pulse energies: $E_P = 20$ and 30 μJ [Fig. 4]. The photon rates in the $W_0$, $W_{-1}$ and $W_{-2}$ bands are plotted in Figs. 4(a&c) and those in the mixing signals in Figs. 4(b&d). Note that, for clarity, the sum of the rates in the two up-shifted sidebands $M_1$ at 730 nm (411 THz) and $M_2$ at 560 nm (535 THz) is plotted, along with the signal $M_0$ at 1048 nm (286 THz); the full experimental data is available in Fig. S2 of Supplement 1.

In Fig. 4(a) the pump pulse energy was $E_P = 20$ μJ (injected into the kagomé-PCF with ~40% coupling efficiency). As the pressure increases, $W_0$ converts to $W_{-1}$ via SRS, giving rise to a $C_w$ and becoming significantly depleted. At 12 bar $W_{-1}$ is itself depleted and light appears at the $W_{-2}$ frequency (i.e., the second Stokes). This is because at this pressure both the $W_0 \to W_{-1}$ and $W_{-1} \to W_{-2}$ processes are phase-matched to the same $C_w$. At higher pressures the $W_{-1} \to W_{-2}$ phase-mismatch increases and the $W_{-2}$ signal drops again.

At 14.2 bar, the $C_w$ created in the $W_0 \to W_{-1}$ process is able to phase-match $M_0 \to M_1$ conversion from 1048 to 730 nm, with photon number conversion efficiencies (based on depletion of the $M_0$ signal) well above 50% at the peak [Fig. 4(b)].

At 30 μJ pump pulse energy, a strong $W_{-2}$ signal appears for pressures above 15 bar [Fig. 4(c)]. This is because the $W_{-1}$ signal is strong enough to reach the threshold for SRS, generating a strong $W_{-2}$ signal and an independent $C_w$. This new $C_w$ causes a new phase-matching point to appear at a pressure of ~27 bar, resulting in a second peak of $M_0 \to M_1$ conversion [Fig. 4(d)]. At the same time the first conversion peak reaches a conversion efficiency above 70%.

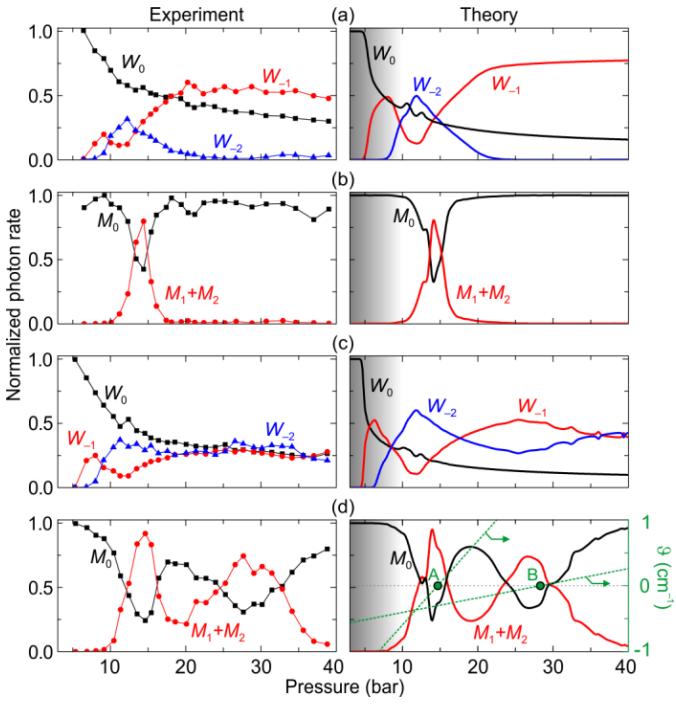

Fig. 4. Experimental and theoretical normalized photon rates of the various signals, plotted against increasing pressure. (a) Pump ($W_0$), first ($W_{-1}$) and second ($W_{-2}$) Stokes waves for $E_P = 20$ μJ. (b) Mixing ($M_0$) and up-shifted anti-Stokes ($M_1 + M_2$) signals, coupled by the $C_w$'s created in (a). (c) Pump ($W_0$), first ($W_{-1}$) and second ($W_{-2}$) Stokes waves for $E_P = 30$ μJ. (d) Mixing ($M_0$) and up-shifted anti-Stokes ($M_1 + M_2$) signals, coupled by the $C_w$'s created in (c). The green lines show the pressure-dependence of the phase-match parameter $\vartheta$. Note that the $C_w$ generated by $W_{-1} \to W_{-2}$ conversion creates a second phase-matching pressure at point B. Up-shifting to $M_1$ is most efficient at the phase-matching pressures (points A and B).

As seen in the right-hand column of Fig. 4, these measurements are in good overall agreement with numerical solutions of a set of coupled spatio-temporal Maxwell-Bloch equations [13]. Note that the model is only valid for pressures above 10 bar, i.e., outside the shaded region in Fig. 4, when the Raman gain is independent of pressure (see Supplement 1 for details).

The simulations also allow us to study the behavior of the $C_w$ amplitude $|Q|$, a quantity that is not accessible in the experiment. Figure 5 plots the evolution of both $|Q|$ and the ($M_1 + M_2$) photon rate along the fiber at gas pressures of 12 and 27 bar for $E_P = 30$ μJ. At 12 bar $|Q|$ grows through SRS-related exponential gain in the $W_{-1}$ signal [upper left-hand panel in Fig. 5]. It peaks at ~40 cm, decreasing thereafter because by that point the majority of pump photons have been converted to the $W_{-1}$ band. Note that the temporal peak of $|Q|$ is delayed by ~1 ns relative to the center of the pump pulse, as expected in transient Raman scattering. For this special pressure (as explained above in Fig. 2) the same $C_w$ is simultaneously able to seed down-conversion from the $W_{-1}$ to the $W_{-2}$ band (and thus be amplified), giving rise to the strong $W_{-2}$ signals observed in Figs. 4(a&c). As expected, up-conversion from the $M_0$ to the $M_1$ band is highest when $|Q|$ is strongest [upper right-hand panel in Fig. 5].

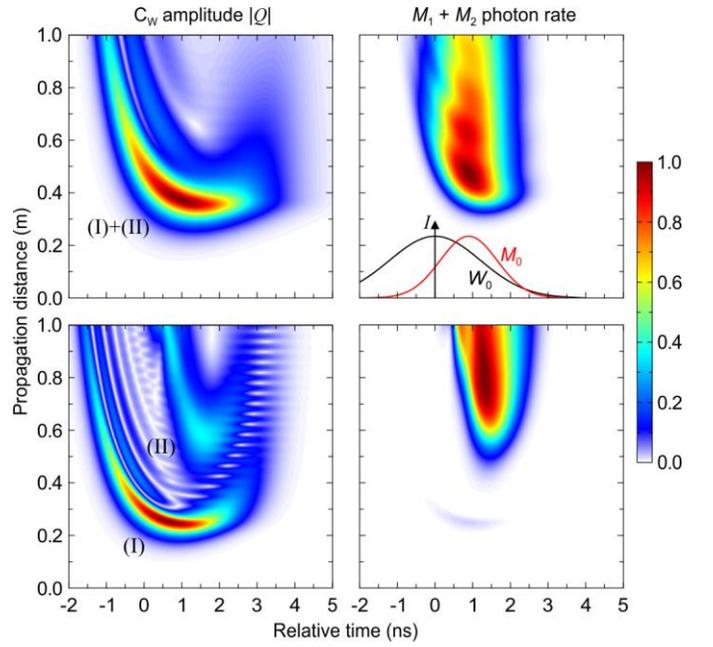

Fig. 5. Simulated spatio-temporal evolution of the normalized $C_w$ amplitude $|Q|$ and normalized ($M_1 + M_2$) photon rate for $E_P = 30$ μJ at pressures of 12 bar (upper) and 27 bar (lower). $C_w$'s created in the $W_0 \to W_{-1}$ process are marked with (I) and those created by the $W_{-1} \to W_{-2}$ process with (II). The inset in the upper right panel shows the normalized temporal profiles of the $W_0$ and $M_0$ pulses at entrance to the fiber. Time is relative to a frame traveling at the group velocity of the $W_0$ pulse. The origin of the periodic oscillations in the lower left panel is discussed in Supplement 1.

This special dual phase-matching condition is no longer valid at 27 bar, because the $W_0 \to W_{-1}$ and $W_{-1} \to W_{-2}$ conversion processes produce different $C_w$'s. As a result, the first peak in $|Q|$ at ~25 cm originates from $W_0 \to W_{-1}$, whereas the second peak at ~60 cm is created by $W_{-1} \to W_{-2}$ conversion and appears at a later time [lower left-hand panel in Fig. 5]. For the mixing beams, although $M_0 \to M_1$ conversion is very weak at the 25 cm point due to strong dephasing ($\vartheta = 1.6$ cm$^{-1}$), it is strong at 60 cm because of phase-matching with the second $C_w$ (lower right-hand panel in Fig. 5). This illustrates the supreme importance of phase-matching for efficient frequency conversion.

From coupled mode theory, the bandwidth $\Delta \nu$ of the up-conversion process covers the range $-\pi < \vartheta L_c < \pi$ where $L_c$ is the length over which the $C_w$ amplitude is significant (~10 cm in the experiments). $\Delta \nu$ depends on the curvature of the dispersion curve for the $M_0 \to M_1$ process and ranges from ~250 THz for $M_0$ at 800 THz, to ~20 THz for $M_0$ at 300 THz. These large bandwidths make it possible to frequency-shift and replicate broadband signals, as seen in Fig. 6 where the full recorded spectrum at 14.5 and 27.6 bar is plotted for $E_P = 30$ μJ. Cascaded up-shifting to the $M_2$ band is possible in the case of 14.5 bar [upper panel in Fig. 6] because of the relatively flat dispersion curve from ~300 to ~600 THz [see for instance the 12 bar curve in Fig. 2(a)]. In contrast, at 27.6 bar efficient conversion is only possible to the first ($M_1$) sideband (lower panel in Fig. 6), owing to the strong curvature of the dispersion over the whole frequency range [see for instance the 30 bar curve in Fig. 2(a)].

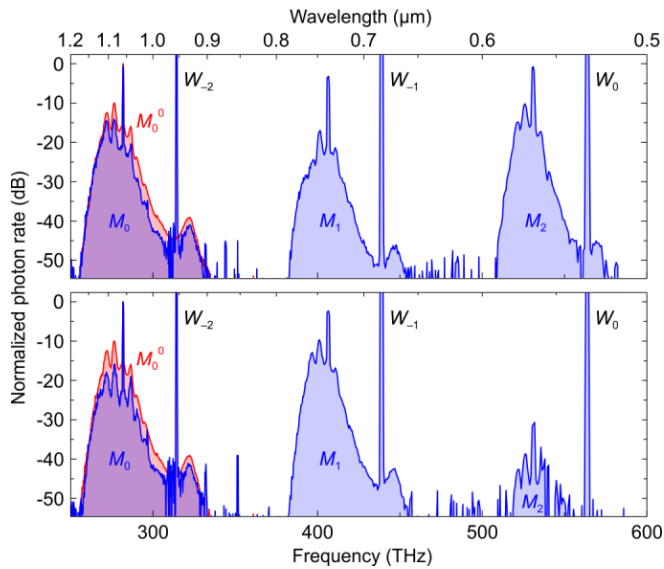

Fig. 6. Demonstration of broadband frequency shifting for 14.5 bar (upper panel) and 27.6 bar (lower panel). The broadband mixing signal $M_0$ is frequency up-shifted to the first ($M_1$) and second ($M_2$) anti-Stokes bands. The original spectrum $M_0^0$ (pump switched off) is depleted accordingly. The photon rates are normalized to the peak of the broadband mixing $M_0^0$ signal at 1064 nm.

Since $\vartheta$ is also a function of pressure, it determines the width of the conversion peaks as seen in Fig. 4(d), where $\vartheta$ is plotted for the two $C_w$'s excited in that case. It is clear that a weaker pressure dependence of $\vartheta$ translates into broader conversion peaks.

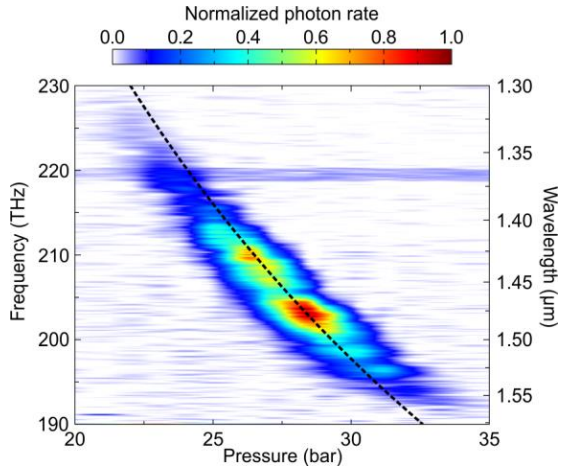

Fig. 7. Frequency down-shifted signal versus pressure. The dashed line represents the theoretical perfectly phase-matched frequencies.

In this Letter we have concentrated on frequency up-shifting. As mentioned in the introduction, it is also possible to use a $C_w$ to frequency down-shift a mixing signal. In this case the $C_w$ acts as a (potentially very strong) seed, being amplified in the conversion process. In Fig. 7 we show the result of an experiment with the same setup where a broadband signal at ~325 THz (920 nm) is down-shifted. The maximum conversion efficiency achieved is ~25%. Note also that, since the phase-matching frequency shifts with pressure, so does the peak of the down-shifted frequency band. The experimental trace nicely follows the analytical curve for perfect phase-matching [the dashed curve in Fig. 7]. Taken together with all-$LP_{01}$ operation, this further demonstrates the wide usable frequency range and versatility of this unique wavelength convertor.

In conclusion, the S-shaped dispersion curve in the vicinity of the ZDP makes hydrogen-filled kagomé-PCF an ideal vehicle for highly efficient wavelength conversion of broad-band optical signals. A coherence wave (which behaves like a travelling refractive index grating) is first excited by SRS on one side of the ZDP. It is then used for phase-matched conversion of a signal on the opposite side of the ZDP. The system is widely pressure tunable, permitting highly efficient up and down-conversion, by 125 THz, of $LP_{01}$ signals from the UV to the near-IR. The system offers a new route for frequency conversion of ultrashort laser pulses from mode-locked lasers. Further flexibility is possible by changing the fiber design, working with different Raman active gases, adding buffer gases or working with a different pump wavelength. This uniquely flexible system has many potential applications in ultrasensitive Raman spectroscopy and laser science.

See Supplement 1 for supporting content.

# Broadband-tunable LP$_{01}$ mode frequency shifting by Raman coherence waves in H$_2$-filled hollow-core PCF: supplementary material 1

S. T. BAUERSCHMIDT,* D. NOVOA, A. ABDOLVAND, AND P. ST.J. RUSSELL

*Max Planck Institute for the Science of Light, Guenther-Scharowsky Strasse 1, 91058 Erlangen, Germany*

*Corresponding author: sebastian.bauerschmidt@mpl.mpg.de*

**We describe the theoretical model employed to simulate the dynamics of the system and show the individual photon rates of the first two up-shifted sidebands ($M_1$ and $M_2$), where only their sum was shown in the main text. In addition, we also provide details on the pressure dependence of the interference of two coherence waves.**

## 1. Numerical simulations

To simulate the spatio-temporal dynamics of the C$_w$ amplitude $Q$ and the optical fields, we used a semi-classical treatment of stimulated Raman scattering [1] based on the following set of coupled Maxwell-Bloch equations:

$$\frac{\partial}{\partial z} E_l = -i\kappa_{2,l-1} Q E_{l-1} q_{l-1} q_l^* \frac{\omega_l}{\omega_{l-1}} \\ -i\kappa_{2,l} Q^* E_{l+1} q_{l+1} q_l^* - \frac{1}{2}\alpha_l E_l \quad \textbf{(S1)}$$

$$\frac{\partial}{\partial t} Q = -Q/T_2 + in\frac{1}{4}\sum_l \kappa_{1,l} E_l^* E_{l+1} q_l^* q_{l+1}, \quad \textbf{(S2)}$$

where the integer index $l$ denotes the Raman sideband of the pump at frequency $\omega_l = \omega_W + l\,\Omega_R$ ($\omega_W$ is the frequency of the pump). The complex electric field amplitudes $e_l(z,\tau)$ are described in terms of the slowly varying envelopes $E_l(z,\tau)$:

$$e_l(z,\tau) = E_l(z,\tau) q_l = E_l(z,\tau) e^{i\beta_l z} \quad \textbf{(S3)}$$

where the propagation constants $\beta_l$ are calculated using Eq. 2 in the main text. The parameters $\alpha_l$ are the attenuation constants of the different spectral lines. We assume the population difference is $n = -1$, i.e., the majority of the molecules remain in the ground state, which is a reasonable approximation since the highest value of coherence $|Q|$ reached in the experiments is ∼0.01. The frequency-dependent coupling coefficients $\kappa_{1,l}$ and $\kappa_{2,l}$ can be derived from the experimentally determined gain values $\gamma$ [2] by:

$$\kappa_{1,l} = \sqrt{\frac{2\gamma c^2 \varepsilon_0^2}{N T_2 \hbar \omega_l}} \quad \textbf{(S4)}$$

and

$$\kappa_{2,l} = \frac{N\hbar \omega_l \kappa_{1,l}}{2\varepsilon_0 c}, \quad \textbf{(S5)}$$

where $N$ is the molecular number density, $c$ is the speed of light in vacuum, $\varepsilon_0$ is the vacuum permittivity, $h$ is Planck's constant and $T_2$ is the dephasing time of the molecules, given in terms of the linewidth $\Delta\nu \sim 1/T_2\pi$ of the Raman transition [2,3].

To model the dynamics of the mixing beam fields $E_m$ ($\omega_m = \omega_M + m\Omega_R$), where $m$ is an integer) we used the same set of coupled field equations as for the pump (Eq. S1), except that $Q$ takes the value set by SRS in the preparation stage. Since the mixing fields are weak, they are assumed to have no effect on the C$_w$ amplitude (for stronger mixing fields significant amplification or attenuation of $Q$ will occur). Also, since the pulse lengths are comparable to the fiber length (1 m) group velocity walk-off between the different spectral lines is negligible.

The first Stokes signal is generated from noise in the experiment, initiating all the observed dynamics. In the simulations we assume a uniform noise floor of $E_{noise}$ = 500 V/m for all the Raman lines except $l$ = 0. We checked the validity of this assumption by measuring the Stokes intensity as a function of pump intensity and extrapolating the exponential growth of the Stokes signal for low pump intensities. This yielded a value close to $E_{noise}$ = 500 V/m for zero pump signal.

The initial shape of the $W_1$ pulse was assumed to be Gaussian and the modal intensity distribution was modeled by a uniform transverse profile with an effective mode area of $A_{eff}$ = 1.5$a^2$ [4], where $a$ is the core radius in the kagomé-PCF. Finally, we considered the same pump energies in both simulations and experiments.

The attenuation coefficients at the wavelengths used in the experiments were extracted from the loss measurement shown in Fig. S1.

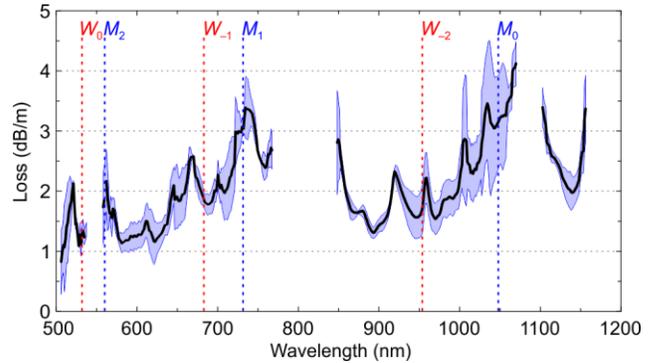

Fig. S1. Measured loss of the HC-PCF (solid line) with estimated error (shaded region). Wavelength ranges without data have loss > 5 dB/m or no loss data are available. The frequency bands used in the Letter are highlighted by the vertical dashed lines.

## 2. Separate photon rates of up-shifted sidebands

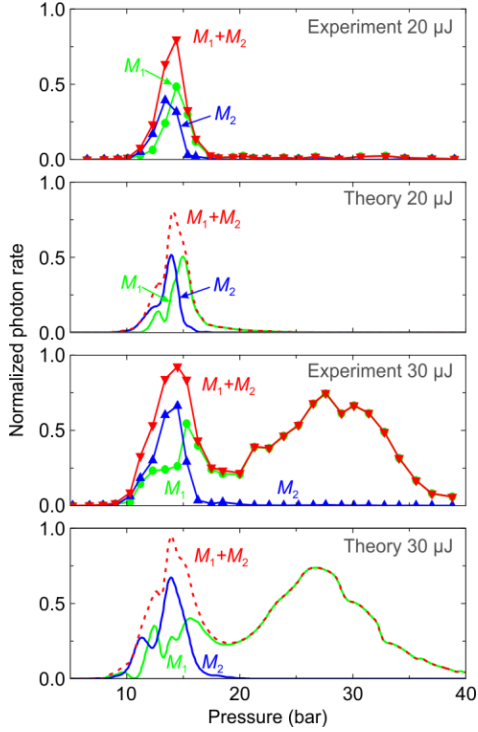

Fig. S2. Experimental and theoretical normalized photon rates of the first ($M_1$) and second ($M_2$) mixing sidebands, plotted against increasing pressure (normalized to the maximum of the $M_0$ rate, not shown) for pump pulse energies $E_P$ of 20 and 30 µJ. Also the sum of both signals ($M_1 + M_2$) is included which was shown in Fig. 3.

In the Letter we restrict the discussion on the dynamics of the mixing beam to the $M_0 \rightarrow M_1$ conversion process, although subsequent scattering to the second sideband ($M_1 \rightarrow M_2$) is also possible (see also Fig. 7 in the main text). Therefore we plot in Fig. 4 of the main text only the sum of the two upshifted signals ($M_1 + M_2$) for clarity. In Fig. S2 we show for completeness the photon rates of the two individual signals together with their sum. Due to the relatively flat dispersion at the phase-matching pressure of the $M_0 \rightarrow M_1$ transition (12 bar, see Fig. 2 in the main text) photons can subsequently be up-shifted from the first to the second sideband giving rise to the $M_2$ signal in Fig. S2 at ~ 12 bar. In contrast, at 30 bar the dispersion is no longer flat and scattering to the $M_2$ band is strongly suppressed. No significant $M_3$ signal was observed in our experiments.

## 3. Coherence wave interference

The periodic pattern along $z$ observed in the coherence at 27 bar in Fig. 5 is caused by interference of the two coherence waves excited in the system. The observed beat length $L_B = 3.9$ cm agrees well with the value predicted from the dispersion relation for the light, given by:

$$L_B = \frac{2\pi}{\Delta\beta_{(W_0 \rightarrow W_{-1})} - \Delta\beta_{(W_{-1} \rightarrow W_{-2})}} \,. \qquad \textbf{(S6)}$$

Fig. S3 shows how these interference fringes change with pressure and propagation distance for a pump pulse energy of 30 µJ and $\tau = 2$ ns. At the dual phase-matching pressure of 12 bar, $L_R \rightarrow \infty$ and no fringes are visible along $z$, whereas for increasing pressures spatial oscillations of $|Q|$ are observed. These oscillations decrease in period as the pressure (and therefore the phase-mismatch) increases. In addition, the onset of the $W_{-1} \rightarrow W_{-2}$ coherence wave is also affected by the increasing phase mismatch; as the pressure increases, this $C_w$ moves closer to the fiber endface because coupling from $W_{-1}$ to $W_{-2}$ by the $W_0 \rightarrow W_{-1}$ coherence wave becomes less efficient. Thus, the phase-matched conversion process at 12 bar turns smoothly into to a cascaded process where the second Stokes band is generated via SRS from the first Stokes band. Note that the growth of the $C_w$ amplitude is retarded for pressures below 10 bar because of decreasing gain, which prevents clear identification of the interference fringes.

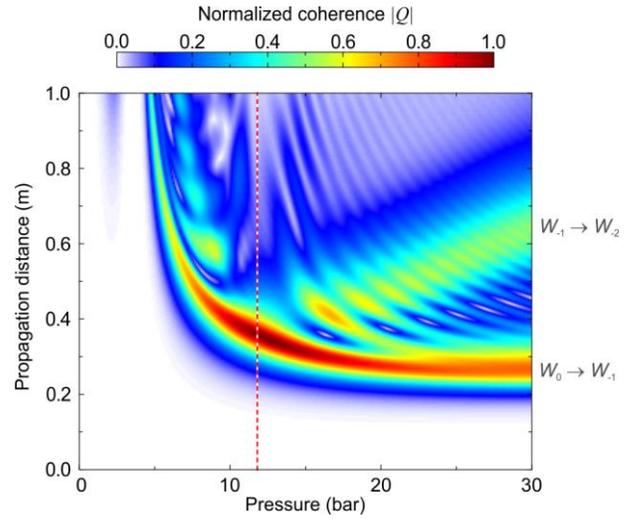

Fig. S3. Amplitude of the coherence $|Q|$ at fixed time $\tau = 2$ ns as a function of pressure and propagation distance. The dashed vertical line marks the pressure where the two $C_w$'s are identical.